\documentstyle[preprint,aps]{revtex}
\tighten

\begin{document}
\draft
\preprint{\vbox{\hfill ADP-93-219/T137\\
                \null\hfill hep-ph/9405306}}
\title{Relativistic Quark-Antiquark Bound State Problem with\\
       Spin-dependent Interactions in Momentum Space}
\author{H.-C.\ Jean and D.\ Robson}
\address{Department of Physics, Florida State University,
         Tallahassee, FL 32306}
\author{A.\ G.\ Williams\cite{byline}}
\address{Department of Physics and Supercomputer Computations Research
         Institute, Florida State University, Tallahassee, FL 32306}
\date{\today}
\maketitle

\begin{abstract}
The work described in this paper is the first step toward a
relativistic three-quark bound-state calculation using a Hamiltonian
consistent with the Wigner-Bargmann theorem and macroscopic locality.
We give an explicit demonstration that we can solve the two-body
problem in momentum space with spin-dependent interactions. The form
of the potential is a combination of linear+Coulomb+spin-spin+%
spin-orbit+tensor, which includes confinement and is of the general
form consistent with rotation, space-reflection and time-reversal
invariance. Comparison is made with previous calculations using an
alternate technique and with the experimental meson mass spectrum. The
results obtained suggest that the model is realistic enough to provide
a two-body basis for the three quark baryon problem in which the
Poincar\'{e} group representation is unitary and cluster separability
is respected.
\end{abstract}

\narrowtext

\section{Introduction}

The work described in this paper is the first step toward a
relativistic three-fermion calculation. For light quark systems,
wherein mesons are described by quark-antiquark dynamics and baryons
by three quark dynamics, one calculable approach is the relativistic
potential model. Covariant approaches based on the Bethe-Salpeter
equation\cite{salpeter} appear to require some form of reduction to a
soluble equation, e.g., Salpeter's instantaneous
approximation\cite{salpeter2}, ladder approximations in Euclidean
space\cite{munczek} or a relativistic equation in which one of the
quarks is restricted to its mass shell\cite{gross}. Each of these
reductions can be argued to have its own strengths and weaknesses.
There is currently no truly satisfactory solution of the
Bethe-Salpeter equation for bound states of two fermions, let alone
three. Therefore, we resort to a potential model with a model
Hamiltonian consistent with the principle of relativity and
macroscopic locality\cite{keister} to solve the three light-quark
bound-state problem. The three-body mass operator (Hamiltonian) can be
defined using the Bakamjian-Thomas construction\cite{bakamjian}.
Unfortunately such a relativistic Hamiltonian is already very
difficult to solve\cite{keister,jean}. One feasible approach is to
solve the Faddeev equations in momentum space using appropriate
two-body wave functions as a basis\cite{jean,glockle}. This
necessarily requires an accurate calculation of such a two-body
wave function basis. We have chosen the collocation method to solve
the two-body equation rather than variational techniques for this
reason. Consequently, it is necessary to first solve the
quark-antiquark problem accurately in momentum space (because of the
difficulty arising from the square root operator in coordinate space
using the collocation method) with general spin-dependent
interactions. This has not previously been done. Previous work in
momentum space has been limited to spin-independent
calculation\cite{jean,maung}. In the present work we attempt to ensure
that the two-body basis is optimal, we fit the meson masses (i.e.,
quark-antiquark systems) with a parametrized interaction in momentum
space including a full range of spin-dependent interactions.

In section II we describe the relativistic two-body Hamiltonian in
terms of a linear confinement term , a Coulomb term, and various
spin-dependent pieces.  We emphasize that this is not a simple
nonrelativistic reduction of an effective one-gluon-exchange potential
since the coefficients of the various terms in the potential will be
constrained by phenomenological considerations alone. The various
terms making up the potential are used because they are invariant
under rotations, space-reflection and time-reversal. To treat
singularities in the spin-dependent interactions a form factor is
introduced for the quark-gluon vertex. Since the nonperturbative
quark-gluon vertex is in general momentum-dependent, this is a natural
thing to include. Numerical solutions are obtained for the model of
reference\cite{carlson2} using partial wave expansions of the various
interactions given in section II. Results for various parameter
choices including those of reference\cite{carlson2} are presented in
section III. These results show that the light meson spectrum can be
well described by such a model. Planned future directions involving
other properties of the meson system, such as electromagnetic form
factors, and the extension to relativistic three-fermion systems are
given in section IV.

\section{Formalism}

In our relativistic potential model, the Hamiltonian $H$ is the sum of
a relativistic kinetic energy operator $T$ and a potential operator
$V$. The kinetic energy operator has the form
\begin{equation}
T = \sqrt{M_1^2+{\bf k}^2} + \sqrt{M_2^2+{\bf k}^2} ~,
\end{equation}
where $M_1$, $M_2$ are the masses of the two quarks and ${\bf k}$ is
the relative momentum. In what follows, we assume $M\equiv M_1=M_2$.
The potential is the sum of a linear confining potential and a short
range spin-dependent interaction:
\begin{equation}
V = V_L + V_{S} ~.
\end{equation}
Here we use a linear confining potential $V_L=\sqrt{\sigma}r$ where
$\sqrt{\sigma}$ is the string tension and $r$ is the relative
coordinate. As a guide to choosing an effective spin-dependent short
range interaction we begin by noting that the nonrelativistic
reduction of the one-gluon-exchange potential in momentum space has
the form\cite{machleidt}
\widetext
\begin{eqnarray}
\langle {\bf k}|V_{NR}|{\bf k}' \rangle &=&
 f_c \alpha_s {1\over2\pi^2} \Bigg\{ {1\over({\bf k}'-{\bf k})^2}
 - {1\over 6M^2} (\mbox{\boldmath$\sigma$}_1\cdot
 \mbox{\boldmath$\sigma$}_2)
 + {3\over 4M^2} {1\over({\bf k}'-{\bf k})^2}
 i (\mbox{\boldmath$\sigma$}_1 + \mbox{\boldmath$\sigma$}_2)
 \cdot{\bf k}\times{\bf k}' \nonumber\\
&& + {1\over4M^2} \left[ (\mbox{\boldmath$\sigma$}_1\cdot\hat{\bf q})
 (\mbox{\boldmath$\sigma$}_2\cdot\hat{\bf q})
 - {1\over3}
 (\mbox{\boldmath$\sigma$}_1\cdot\mbox{\boldmath$\sigma$}_2)
 \right]\Bigg\},\label{eq.gluonp}
\end{eqnarray}
\narrowtext
where $\alpha_s$ is the strong-interaction fine-structure constant,
$f_c$ is the color factor (which is $-4/3$ for quark-antiquark and
$-2/3$ for quark-quark), $\mbox{\boldmath$\sigma$}_1$,
$\mbox{\boldmath$\sigma$}_2$ are the Pauli matrices and
${\bf q}\equiv{\bf k}'-{\bf k}$ is the momentum transfer. It is
readily seen that the four terms in the curly brackets represent the
Coulomb potential, the spin-spin, spin-orbit and tensor interactions
respectively. These are easily recognized in coordinate space
representation:
\widetext
\begin{eqnarray}
\langle {\bf r}|V_{NR}|{\bf r}' \rangle &=& \delta({\bf r}-{\bf r}')
 f_c \alpha_s \Bigg\{ {1\over r} - {2\pi\over 3M^2} \delta({\bf r})
 (\mbox{\boldmath$\sigma$}_1\cdot\mbox{\boldmath$\sigma$}_2)
 - {3\over 2M^2 r^3} {\bf L}\cdot{\bf S} \nonumber\\
&& - {3\over4M^2 r^3} \left[
 (\mbox{\boldmath$\sigma$}_1\cdot\hat{\bf r})
 (\mbox{\boldmath$\sigma$}_2\cdot\hat{\bf r}) - {1\over3}
 (\mbox{\boldmath$\sigma$}_1\cdot\mbox{\boldmath$\sigma$}_2)
 \right]\Bigg\},\label{eq.gluonr}
\end{eqnarray}
\narrowtext
where ${\bf S}={1\over2}(\mbox{\boldmath$\sigma$}_1 +
\mbox{\boldmath$\sigma$}_2)$ is the total spin operator and {\bf L} is
the orbital angular momentum operator. The spin-spin interaction
includes a delta function of {\bf r}, (i.e., it is a contact
interaction), and so we introduce a Gaussian form factor
$\exp(-{1\over2}\Lambda^2 {\bf q}^2)$ at the quark-gluon vertex as in
Ref.\ \cite{carlson2}. The variable $\Lambda$ can be interpreted as
the size of the quark. In Ref.\ \cite{stanley} a form factor
$1 / ({\bf q}^2+\beta^2)$ in which $\beta^{-1}$ is the effective quark
size is used to eliminate the singularity. Since the
one-gluon-exchange potential (Eq.\ \ref{eq.gluonp}) is derived via a
nonrelativistic reduction (and so can not represent a reasonable
interaction in a relativistic calculation) and in any case is really
an effective interaction, we are completely free to vary the relative
strengths of these interactions. In particular it is well known that
the phenomenological strength of the spin-orbit interaction is much
weaker than that predicted by the nonrelativistic reduction of the
one-gluon-exchange potential\cite{godfrey}. Hence we use multiplying
factors $C_L, C_C, C_{SS}, C_{LS}, C_T$ so that the strength of every
term can be varied to fit the meson data. The general form of the
potential that we used in our calculation then has the form
\begin{eqnarray}
\langle {\bf k}|V|{\bf k}' \rangle &=& C_L V_L +
 {f_c \alpha_s \over2\pi^2} e^{-\Lambda^2 {\bf q}^2}
 \Bigg\{ {C_C\over({\bf k}'-{\bf k})^2} \nonumber\\
&& - {C_{SS}\over 6M^2}
 (\mbox{\boldmath$\sigma$}_1\cdot\mbox{\boldmath$\sigma$}_2)
 \nonumber\\
&& + {3C_{LS}\over 4M^2} {1\over({\bf k}'-{\bf k})^2}
 i (\mbox{\boldmath$\sigma$}_1 + \mbox{\boldmath$\sigma$}_2)
 \cdot{\bf k}\times{\bf k}' \nonumber\\
&&+ {C_T\over4M^2} \left[ (\mbox{\boldmath$\sigma$}_1\cdot\hat{\bf q})
 (\mbox{\boldmath$\sigma$}_2\cdot\hat{\bf q}) - {1\over3}
 (\mbox{\boldmath$\sigma$}_1\cdot\mbox{\boldmath$\sigma$}_2)
 \right]\hspace{-1.3pt}\Bigg\}.\label{eq.poten}
\end{eqnarray}
Note that the only potentials that respect rotation invariant,
space-reflection and time-reversal which are not included in our
potential are $(\mbox{\boldmath$\sigma$}_1\cdot{\bf L})
(\mbox{\boldmath$\sigma$}_2\cdot{\bf L})$ and
$(\mbox{\boldmath$\sigma$}_1-\mbox{\boldmath$\sigma$}_2)\cdot{\bf L}$.
The latter term becomes significant only if $l\neq0$ and one of the
quarks is much heavier than other.

To do the calculations in momentum space we have to use a partial-wave
expansion for the potentials in Eq.\ (\ref{eq.poten}). The
difficulty with a momentum-space formulation for the linear potential
was solved in Refs.\ \cite{eyre,spence}. The following are the results
for the partial-wave expansion of the potentials in momentum
space.\hfill\break
\noindent For potentials depending on $|{\bf q}|$ only:
\begin{mathletters}
\begin{equation}
\langle {\bf k}|V|{\bf k}' \rangle = \tilde{V}(|{\bf q}|) ~,
\end{equation}
we find (using $\mu\equiv\hat{\bf k}'\cdot\hat{\bf k}$)
\begin{eqnarray}
 \langle klsjm|V|k'l's'j'm' \rangle &=& \delta_{ll'} \delta_{ss'}
 \delta_{jj'} \delta_{mm'} \nonumber\\
&& \times 2\pi \int_{-1}^{1} \tilde{V}(|{\bf q}|) P_{l}(\mu) d\mu ~.
\end{eqnarray}
\end{mathletters}
\noindent For the spin-spin interaction:
\begin{mathletters}
\begin{eqnarray}
\lefteqn{\langle {\bf k}|V|{\bf k}' \rangle = \tilde{V}({\bf q})
 {\bf S}_1\cdot{\bf S}_2 ~,} \\
\lefteqn{\langle klsjm|V|k'l's'j'm'\rangle = \delta_{ll'} \delta_{ss'}
 \delta_{jj'} \delta_{mm'} } \nonumber\\
&& \hspace{30pt} \times {1\over2} [s(s+1)-s_1(s_1+1)-s_2(s_2+1)]
 \nonumber\\
&& \hspace{30pt} \times 2\pi \int_{-1}^{1} \tilde{V}(|{\bf q}|)
 P_{l}(\mu) d\mu ~.
\end{eqnarray}
\end{mathletters}
\noindent For the spin-orbit interaction:
\begin{mathletters}
\begin{eqnarray}
\lefteqn{\langle {\bf k}|V|{\bf k}' \rangle = \tilde{V}(|{\bf q}|)
 i {\bf S} \cdot {\bf k}\times{\bf k}' ~,}\label{eq.ls.a} \\
\lefteqn{\langle klsjm|V|k'l's'j'm' \rangle = \delta_{ll'}
 \delta_{ss'}\delta_{jj'}\delta_{mm'} }\nonumber\\
&& \hspace{30pt} \times {j(j+1)-l(l+1)-s(s+1)\over2} \nonumber\\
&& \hspace{30pt} \times {2\pi kk'\over 2l+1} \int_{-1}^{1}
 \tilde{V}(|{\bf q}|) [P_{l+1}(\mu)-P_{l-1}(\mu)] d\mu ~,
 \label{eq.ls.b}
\end{eqnarray}
\end{mathletters}
\noindent and for the tensor interaction:
\widetext
\begin{mathletters}
\begin{eqnarray}
&& \langle {\bf k}|V|{\bf k}' \rangle = \tilde{V}(|{\bf q}|) q^2
 \left[ 3(\mbox{\boldmath$\sigma$}_1\cdot\hat{\bf q})
 (\mbox{\boldmath$\sigma$}_2\cdot\hat{\bf q})
 - (\mbox{\boldmath$\sigma$}_1\cdot\mbox{\boldmath$\sigma$}_2)
 \right] ~,\label{eq.tensor.a} \\
&& \langle klsjm|V|k'l's'j'm' \rangle = \delta_{jj'} \delta_{mm'}
 \delta_{s1} \delta_{s'1} 4\sqrt{30}\pi
 \left\{ \begin{array}{ccc}j&1&l\\2&l'&1\end{array} \right\}
 \sum_{L=0}^2 \sum_{l''} (-1)^{l'+j+1} (2l''+1) \nonumber\\
&& \hspace{30pt} \times \sqrt{(2l+1)(2l'+1)(5-2L)C^5_{2L}}
 \left\{\begin{array}{ccc}l'&2\hspace{-2pt}-\hspace{-2pt}L&l''
 \\L&l&2\end{array}\right\} \nonumber\\
&& \hspace{30pt} \times
 \left(\begin{array}{ccc}l''&L&l\\0&0&0\end{array}\right)
 \left(\begin{array}{ccc}l''&2\hspace{-2pt}-\hspace{-2pt}L&l'\\0&0&0
 \end{array}\right)
 k^L k'^{2-L} \int \tilde{V}(|{\bf q}|)P_{l''}(\mu)d\mu \nonumber\\
&& \makebox[30pt][r]{=} \left\{\begin{array}{l}
   \delta_{jj'}\delta_{mm'}\delta_{s1}\delta_{s'1}
   {8l(l+1)-3[l(l+1)-j(j+1)+2][l(l+1)-j(j+1)+1]\over(2l-1)(2l+3)}
   2\pi\\
   \times \int_{-1}^1 \tilde{V}(q) [(k^2+k'^2)P_l(\mu)-kk'
   {2l+3\over2l+1}P_{l-1}(\mu)-kk'{2l-1\over2l+1}P_{l+1}(\mu)] d\mu
   \mbox{~~~if~} l'=l \\
   \delta_{jj'}\delta_{mm'}\delta_{s1}\delta_{s'1}
   {6\sqrt{(L+1)(L+2)}\over(2L+3)} 2\pi\\
   \times \int_{-1}^1 \tilde{V}(q) [k'^2P_{L}(\mu)+k^2P_{L+2}(\mu)-
   2kk'P_{L+1}(\mu)] d\mu \mbox{~~~if~} l'\neq l, L=\mbox{min}(l,l')
 \end{array}\right\} ~.\label{eq.tensor.b}
\end{eqnarray}
\end{mathletters}
\narrowtext
In the above equations $C^n_k$ is a binomial coefficient, $P_l(\mu)$
is a Legendre polynomial,
$\left(\begin{array}{ccc}j_1&j_2&j_3\\m_1&m_2&m_3\end{array}\right)$
is a 3-j symbol and
$\left\{\begin{array}{ccc}j_1&j_2&j_3\\j_4&j_5&j_6\end{array}\right\}$
is a 6-j symbol\cite{edmonds}. Note that equations (\ref{eq.ls.b}) and
(\ref{eq.tensor.b}) are new results. With these two equations, we can
do calculations with spin-orbit and tensor interactions for almost any
functional form of $\tilde{V}(|{\bf q}|)$ in equations (\ref{eq.ls.a})
and (\ref{eq.tensor.a}).

To solve the relativistic wave equation
\begin{eqnarray}
&& 2 \sqrt{M^2+{\bf k}^2} \Psi_{lsj}(k) \nonumber\\
&& + \sum_{l's'} \int_0^\infty
 \langle klsjm|V|k'l's'jm \rangle \Psi_{l's'j}(k') k'^2dk' \nonumber\\
&& \hspace{30pt} = E \Psi_{lsj}(k) \label{eq.h}
\end{eqnarray}
we first introduce an auxiliary function $\psi_{lsj}(k)$ given by
\begin{equation}
\Psi_{lsj}(k) = k^{l-1} \psi_{lsj}(k) ~.
\end{equation}
The power $l-1$ follows from considering the behavior of
$\Psi_{lsj}(k)$ as $k\to 0$. The auxiliary wave function is expanded
in a complete set of basis functions. In Refs.~\cite{stanley} and
\cite{godfrey} the wave functions were expanded in a harmonic
oscillator basis. In Ref.~\cite{brayshaw} the wave functions were
expanded in confinement eigenstates which are solutions of a
relativistic Schr\"odinger equation with a confinement potential only.
Here we choose cubic Hermite splines\cite{prenter} as the basis
functions. The cubic Hermite splines are piecewise polynomials of
degree three with continuous first derivatives\cite{prenter}. Also we
impose the boundary conditions by requiring the auxiliary wave
function be zero at $k=0$ and at a cutoff $k=k_{max}$. By expanding
the wave function in N splines functions and requiring that the
expansion satisfies Eq.~(\ref{eq.h}) at N distinct values (collocation
points) of $k$ Eq.\ (\ref{eq.h}) can be converted to a matrix
equation. The two-point Gaussian quadrature points on each interval
are chosen as the collocation points. The eigenvalues and eigenvectors
of the matrix equation are the masses of the mesons and the
coefficients of the meson wave function bases respectively.

\section{Results}
The light meson mass spectra can be obtained by solving for the
eigenvalues of the relativistic Hamiltonian $H=T+V-M_{q\bar{q}}$ where
V is given in Eq.\ (\ref{eq.poten}) and $M_{q\bar{q}}$ is a constant
energy to be determined from data. If $M_{q\bar{q}}=0$, the model
gives the meson mass spectra; if $M_{q\bar{q}}\neq 0$, then the model
only gives the splittings. In our opinion, this overall constant
energy is somewhat artificial. In our first calculation, the
parameters used are $M=360$ MeV, $f_c\alpha_s=-0.5$ and
$\sqrt{\sigma}=0.197$ (GeV)$^2$. These parameters are chosen so that
the results can be compared with the results of Ref.\ \cite{carlson2}
which used a variational method to solve the wave equation in
coordinate space. Unfortunately, they used a different method to treat
the spin-orbit interaction so we can only directly compare our results
with theirs without spin-orbit interaction, i.e.\ for $\pi, \rho$ and
$b$ mesons. We initially set $C_L=C_C=C_{SS}=1$, $C_{LS}=C_{T}=0$ and
eliminate the vertex form factor for the Coulomb potential. These
results are listed in Table \ref{tbl.1} with the same constant energy
($M_{q\bar{q}}=750$ MeV) subtracted out.

Our results are consistently higher than their results by $25-60$
MeV\cite{carlson1}. We have been unable to establish the origin of the
discrepancy but note that our Fortran codes were cross checked with a
totally different method\cite{parramore} and we agree within numerical
error (i.e., $<0.5$ MeV).

The results of fitting light meson spectra by varying two parameters
$\Lambda$ and $C_{LS}$ ($C_L=C_C=C_{SS}=C_T=1$, $M=360$ MeV, no form
factor for the Coulomb potential) are shown in Table \ref{tbl.2}. The
differences between this second calculation and that of
Ref.\ \cite{carlson2} are due to the following factors:
\begin{enumerate}
 \item[a)] We use $\Lambda=0.6426\mbox{ GeV}^{-1}$; they used
       $\Lambda=0.13$ fm $=0.658805\mbox{ GeV}^{-1}$.
 \item[b)] The spin-orbit interaction is treated differently and our
       strength of the spin-orbit interaction is $C_{LS}=0.3236$.
 \item[c)] We consider the coupling between different channels via the
       tensor interaction.
 \item[d)] A different constant energy ($M_{q\bar{q}}=738.24$ MeV) is
       used here versus 750 MeV.
\end{enumerate}
The average deviation from the fifteen experimental values is 69 MeV
for our results and 92 MeV for Ref.\ \cite{carlson2}.

As we mentioned in the previous section, we are free to adjust all
seven parameters $(M$, $\Lambda$, $C_L$, $C_C$ ,$C_{SS}$, $C_{LS}$,
$C_{T})$ to fit the
light meson spectra. We also eliminate the need for the constant
energy term $M_{q\bar{q}}$ used above and use a
form factor not only for spin-spin, spin-orbit and tensor interaction
but also for the Coulomb interaction. The parameters we use are
$M=258$ MeV, $\Lambda=0.645$ GeV$^{-1}$,
$C_L=0.6704,~C_C=3.3824,~C_{SS}=0.155,~C_{LS}=0.0448$ and
$C_{T}=0.1868$. The results of this fitting are listed in the last
column of Table \ref{tbl.2}. The average deviation is 28 MeV for
eleven states which can be compared to the average deviation of 55 MeV
in a similar model calculated in Ref.\ \cite{stanley}. The average
deviation from all fifteen experimental values is 51 MeV which is
significantly better than that obtained by only varying two parameters
(69 MeV).

The above parameter values reflect the following: The linear
confinement potential is a little weaker than for static quarks
($C_L\simeq 1$). The $C_C$ value is considerably greater than one,
which means that light quark systems have a running coupling constant
($\alpha_s C_C$) much larger than heavy quark systems as might be
expected from QCD (asymptotic freedom). $C_{LS}$ is very weak as
observed by others\cite{godfrey}. $C_{SS}$ and $C_T$ are also reduced
relative to values obtained for heavy quarks.

\section{Future Work}

Potential models based on the constituent quark model have enjoyed
considerable success. Even the nonrelativistic potential model works
much better than one might na\"{\i}vely expect. However, the
nonrelativistic constituent quark model fails to explain the meson
and baryon form factors at high energy because the calculated form
factors at high energy fall off too fast. In our calculations, the
tails of the momentum space wave functions have a power-law fall off,
rather than the exponential decay of harmonic oscillator potential
models. It is hoped that using a wave function like ours with an
appropriate relativistic ``boost'' might be able to provide the
correct form factor behavior at high energy. This question is
currently being pursued.

In 1939, Wigner\cite{wigner} showed that the description of a physical
system in a relativistic quantum theory must correspond to a unitary
representation of the Poincar\'{e} group. In 1961, Foldy\cite{foldy}
recognized the importance of macroscopic locality (cluster
separability) as an additional constraint on relativistic potential
models. Many ``relativistic'' three-body potential
models\cite{capstick} do not
satisfy either of these two requirements. Relativistic three-body mass
operators which satisfy both of these two requirements have been
discussed in a recent review\cite{keister}. A relativistic
three-body bound state formulation which incorporates Wigner's theorem
and macroscopic locality can be solved by using the relativistic
Faddeev equation for the eigenvalues and eigenvectors of the
three-body mass operator in momentum space. The calculations are very
complicated since the potential operators are functions of the
position operators and are inside the square roots. These difficulties
can be overcome by using relativistic two-body wave functions as the
basis\cite{jean,glockle}. Therefore, solving the two-body equation
accurately in momentum space was necessary as a first step. The
two-body calculations discussed in this paper are important primarily
because of their use as input for three-quark calculations. Our
ultimate purpose is to solve the relativistic bound state problem for
three light constituent quarks.

\acknowledgments
We are grateful to Joel Parramore for providing assistance with
cross-checking our results and to Jorge Piekarewicz for helpful
discussions. This work was supported by the U.S. Department of Energy
through Contract No. DE-FG05-86ER40273, and by the Florida State
University Supercomputer Computations Research Institute which is
partially funded by the Department of Energy through Contract
No.\ DE-FC05-85ER250000. The work of AGW was also partially supported
by the Australian Research Council.

\narrowtext
\begin{table}
\begin{center}
\begin{tabular}{clrrr}
$nLSJ^\pi$ & ~state & Exp.\ value & $E-750$ & Ref.\ \cite{carlson2}
 \\ \hline
$0000^{-}$ & $\pi$ & 138~~~~ & 164.8 &140\\
$0011^{-}$ & $\rho$ & 768\makebox[0pt][l]{.3}~~~~ & 810.9 &751\\
$0101^{+}$ & $b_1$ & 1233~~~~ & 1155.0 &1113\\
$1000^{-}$ & $\pi(1300)$ & 1300~~~~ & 1143.2 &1114\\
$1011^{-}$ & $\rho(1450)$ & 1450~~~~ & 1490.6 &1442\\
$0202^{-}$ & $\pi_2$ & 1665~~~~ & 1603.5 &1560\\
$0303^{+}$ & $b_3$ & & 1930.6 &1892\\
$2011^{-}$ & $\rho(1700)$ & 1700~~~~ & 1995.4 &1957\\
$0404^{-}$ & $\pi_4$ & & 2214.0 &2177\\
$0505^{-}$ & $b_5$ & & 2470.8 &2438\\
\end{tabular}
\end{center}
\caption{The light meson mass spectrum without spin-orbit and tensor
interactions. Our results (the fourth column) are consistently higher
than those of Ref.\ \protect\cite{carlson2} by 25-60 MeV.}
\label{tbl.1}
\end{table}
\mediumtext
\begin{table}
\begin{center}
\begin{tabular}{clrrrrr}
$nLSJ^\pi$ & ~state & Exp.\ value & Ref.\ \cite{carlson2} & $E-738.24$
 & Ref.\ \cite{stanley} & $E$ \\ \hline
$0000^{-}$ & $\pi$ & 138~~~~ & 140 & 133.6 & 132 & 140.1\\
$0011^{-}$ & $\rho$ & 768\makebox[0pt][l]{.3}~~~~ & 751 & 807.2 & 755
 & 775.7\\
$0101^{+}$ & $b_1$ & 1233~~~~ & 1113 & 1165.0 & 1106 & 1174.6\\
$0110^{+}$ & $a_0$ & 983\makebox[0pt][l]{.3}~~~~ & 931 & 971.2 & 1108
 & 973.7\\
$0111^{+}$ & $a_1$ & 1260~~~~ & 1180 & 1251.7 & 1273 & 1298.2\\
$0112^{+}$ & $a_2$ & 1318\makebox[0pt][l]{.4}~~~~ & 1256 & 1300.2
 & 1328 & 1323.0\\
$1000^{-}$ & $\pi(1300)$ & 1300~~~~ & 1114 & 1140.8 & 1194 & 1188.9\\
$1011^{-}$ & $\rho(1450)$ & 1450~~~~ & 1442 & 1473.3 & 1521 & 1472.7\\
$0202^{-}$ & $\pi_2$ & 1665~~~~ & 1560 & 1615.7 & 1675 & 1661.9\\
$0213^{-}$ & $\rho_3(1690)$ &1691~~~~ &1620 & 1668.0 & 1754 & 1702.2\\
$0313^{+}$ & $a_3(2050)$ & 2050~~~~ & 1894 & 1944.3 & & 1981.6\\
$0314^{+}$ & $a_4(2040)$ & 2040~~~~ & 1933 & 1980.0 & 2100 & 2001.1\\
$2011^{-}$ & $\rho(1700)$ & 1700~~~~ & 1957 & 2004.4 & & 1960.5\\
$0415^{-}$ & $\rho_5(2350)$ & 2350~~~~ & 2207 & 2257.7 & & 2256.2\\
$0516^{+}$ & $a_6(2350)$ & 2450~~~~ & 2462 & 2510.9 & & 2483.1
\end{tabular}
\end{center}
\caption{The light meson mass spectrum by varying two parameters
$\Lambda$ and $C_{LS}$ (the fifth column) and those by varying all 7
parameters (the last column). The average deviation from the fifteen
experiment values is 69 MeV for the second calculation (the fifth
column) and 51 MeV for the third calculation (the last column).}
\label{tbl.2}
\end{table}


\begin{thebibliography}{99}
\bibitem{parramore} This is a nonrelativistic calculation in
        coordinate space with wave functions expanded in a harmonic
        oscillator basis kindly provided by Joel Parramore (private
        communication).
\end{thebibliography}
\end{document}